\documentclass[conference]{IEEEtran}
\usepackage{cite}
\usepackage{graphicx,color,epsfig,rotating}
\usepackage{amsfonts,amsmath,amssymb}
\usepackage{algorithm,algorithmic}
\usepackage{url}
\usepackage{color}
\usepackage{epsfig}
\usepackage{subfigure}

\IEEEoverridecommandlockouts

\makeatletter
\def\ps@headings{%
\def\@oddhead{\mbox{}\scriptsize\rightmark \hfil \thepage}%
\def\@evenhead{\scriptsize\thepage \hfil \leftmark\mbox{}}%
\def\@oddfoot{}%
\def\@evenfoot{}}
\makeatother

\newtheorem{theorem}{Theorem}
\newtheorem{lemma}{Lemma}

\def\ps@headings{%
\def\@oddhead{\mbox{}\scriptsize\rightmark \hfil \thepage}%
\def\@evenhead{\scriptsize\thepage \hfil \leftmark\mbox{}}%
\def\@oddfoot{}%
\def\@evenfoot{}}
\makeatother

\begin{document}

\title{ Wireless Device-to-Device Communications with Distributed Caching
\thanks{This research was supported in part by NSF Career Grant CCF-1055099 and research gifts by Intel and Microsoft Research.}}
\author{Negin Golrezaei,~Alexandros G. Dimakis,~Andreas F.
Molisch\\ Dept. of Electrical Eng.\\
University of Southern California\\
Los Angeles, CA, USA\\
emails:~\{golrezae,dimakis,molisch\}@usc.edu
}
\maketitle
\thispagestyle{empty}
\pagestyle{empty}

\begin{abstract}
We introduce a novel wireless device-to-device (D2D) collaboration architecture that exploits distributed storage of popular content to enable frequency reuse. 
	We identify a fundamental conflict between collaboration distance and interference and show how to optimize the transmission power to 
	maximize frequency reuse. Our analysis depends on the user content request statistics which are modeled by a Zipf distribution. Our main result is a closed form expression of the optimal collaboration distance as a function of the content reuse distribution parameters. We show that if the Zipf exponent of the content reuse distribution is greater than 1, it is possible to have a number of D2D interference-free collaboration pairs that scales linearly in the number of nodes. 
	If the Zipf exponent is smaller than 1, we identify the best possible scaling in the number of D2D collaborating links. Surprisingly, a very simple distributed caching policy achieves the optimal scaling behavior and therefore there is no need to centrally coordinate what each node is caching.

\end{abstract}
\section{Introduction}

Wireless mobile data traffic is expected to increase by a factor of $40$ over the next five years, from the current $93$ Petabytes to $3600$ Petabytes per month in the next five years~\cite{cisco66}. This explosive demand is fueled mainly by mobile video traffic that is expected to increase by a factor of $65$ times, and become the by far dominant source of data traffic. 
Modern smartphones and tablets have significant storage capacity often reaching several gigabytes. Recent breakthroughs in dense NAND flash will make 128GB smartphone memory chips available in the coming months. In this paper we show how to exploit these storage capabilities to significantly reduce wireless capacity bottlenecks.

The central idea in this paper is that, for most types of mobile video traffic, we can replace backhaul connectivity with storage capacity. This is true because of {\em content reuse}, \textit{i.e.}, the fact that popular video files will be requested by a large number of users. Distributed storage enhances the opportunities for user collaboration.

We recently introduced the idea of femtocaching helpers~\cite{femtocaching} \cite{coded_femtocaching}, small base stations with 
a low-bandwidth (possibly wireless) backhaul link and high storage capabilities. 
In this paper we take this architecture one step further: We introduce 
\emph{a device-to-device (D2D) architecture where the mobiles are used as caching storage nodes}. Users can collaborate by caching popular content and utilizing local device-to-device communication when a user in the vicinity requests a popular file. The base station can keep track of the availability of the cached content and direct requests to the most suitable nearby device. Storage allows users to collaborate even when they do not request the same content \textit{at the same time}. 
This is a new dimension in wireless collaboration architectures beyond relaying and cooperative communications.  

\textbf{Our contributions:} In this paper we introduce the novel D2D architecture and formulate some 
theoretical problems that arise. Specifically, we identify a conflict between collaboration distance and interference. We show how to optimize the D2D collaboration distance
and analyze the scaling behavior of D2D benefits. The optimal collaboration distance depends on the 
content request statistics which are modeled by a Zipf distribution. Our main result is a closed form expression of the optimal collaboration distance as a function of the content reuse distribution parameters. 
We show that if the Zipf exponent of the content reuse distribution is greater than $1$, it is possible 
to have a number of D2D interference-free collaboration pairs that scales linearly in the number of nodes. 

If the Zipf exponent is smaller than $1$, we identify the best possible scaling in the number of D2D collaborating links. Surprisingly, a very simple distributed caching policy achieves the optimal scaling behavior and therefore there is no need to centrally coordinate what each node is caching. 

 The remainder of this paper is organized as follows: In Section~\ref{sec:model} we setup the D2D formulation 
and explain the tradeoff between collaboration distance and interference. 
Section~\ref{sec:analysis} contains our two main theorems, the scaling behavior for Zipf exponents 
greater and smaller than $1$. In Section \ref{sec:discussion} we discuss future directions, open problems and conclusions. Finally, in the Appendix we include some interesting technical parts of our proofs. Due to space constraints we omit the complete proofs from this version of the paper.

\section{Model and Setup}

\label{sec:model}
We consider $n$ users distributed uniformly in a unit square and consider this as single cell. The base station (BS) might be aware of the stored files and channel state information of the users and control the D2D communications. 
For simplicity, we neglect inter-cell interference and consider one cell in isolation. We further assume that the D2D communication does not interfere with communication between the BS and users. This assumption is justified if the D2D communications occur in a separate frequency band (\textit{e.g.}, WiFi). For the device-to-device throughput, we henceforth do not need to consider explicitly the BS and its associated communications.

\begin{figure}
\centerline{\includegraphics[width=5cm]{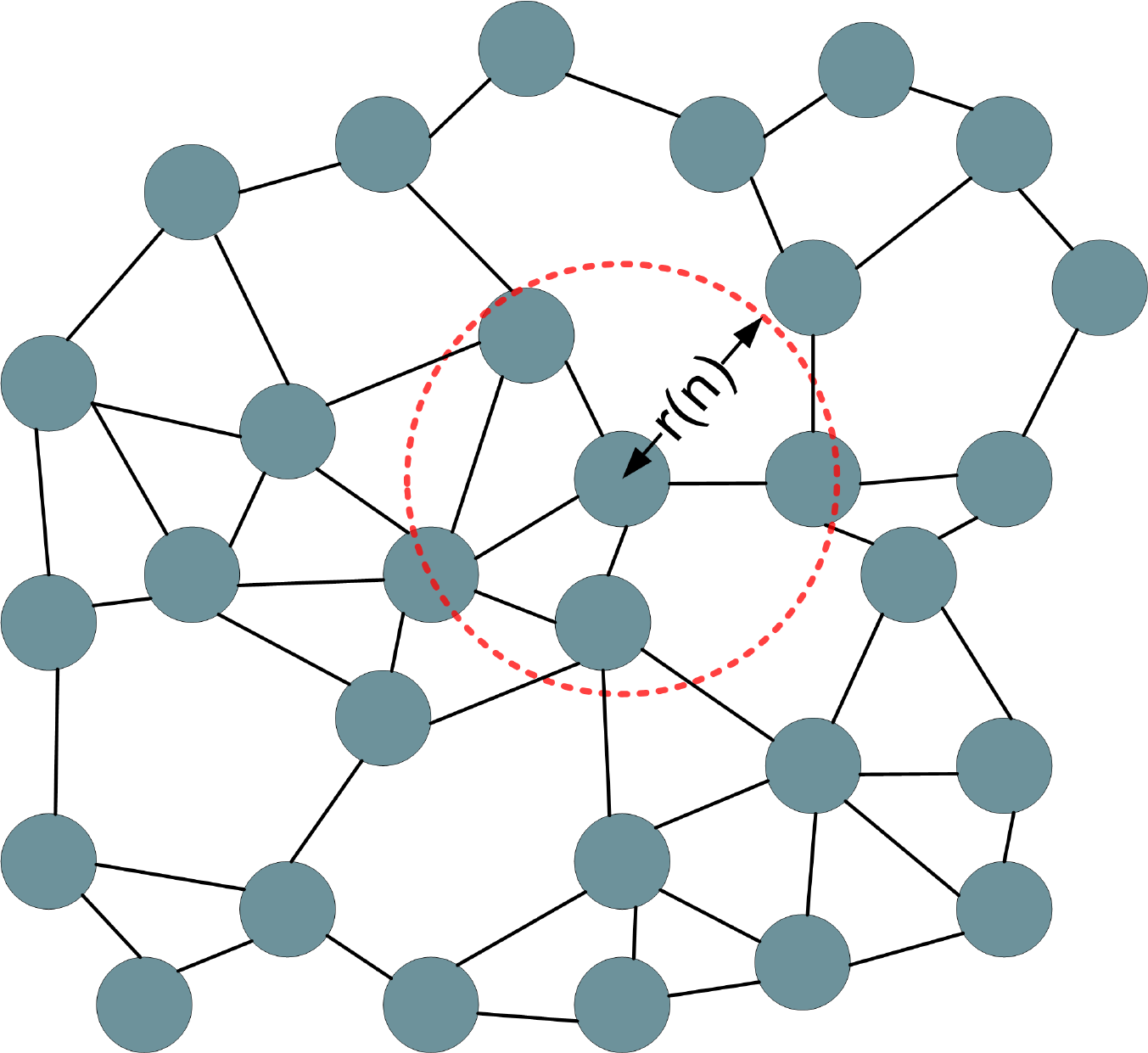}}
\caption{Random geometric graph example with collaboration distance $r(n)$.}
\label{RGG}
\end{figure}

The communication is modeled by random geometric graph $G(n,r(n))$ where two users (assuming D2D communication is possible) can communicate if their physical distance is smaller than some collaboration distance $r(n)$~\cite{gupta2000capacity,RGG book}. The maximum allowable distance for D2D communication $r(n)$ is determined by the  power level for each transmission. Figure \ref{RGG} illustrates an example of random geometric graph (RGG).

We assume that users may request files from a set of size $m$ that we call a ``library''.  
The size of this set should increase as a function of the number of users $n$. Intuitively, the set of YouTube videos requested in Berkeley in one day should be smaller than the set of requested in Los Angeles. 
We assume that this growth should be sublinear in $n$, \textit{e.g.} $m$ could be $\Theta(\log(n))$.

Each user requests a file from the library by sampling independently using a popularity distribution. 
Based on numerous studies, Zipf distributions have been established as good models to the measured popularity of video files \cite{zipf,tracedata}. Under this model, the frequency of the $i$th popular file, denoted by $f_i$, is inversely proportional to its rank:
 \begin{equation}\label{zipf}
 f_i=\frac{{\frac{1}{{{i^{\gamma_r} }}}}}{{\sum\limits_{j = 1}^m {\frac{1}{{{j^{\gamma_r} }}}} }},\,\,\ 1\leq i\leq m.
 \end{equation}
 The Zipf exponent $\gamma_r$ characterizes the distribution by controlling the relative popularity of files.
 Larger $\gamma_r$ exponents correspond to higher content reuse, \textit{i.e.,} the first few popular files account for the majority of requests.

Each user has a storage capacity called cache which is populated with some video files. For our scaling law analysis we assume that all files have the same size, and each user can store one file. This yields a clean formulation and can be easily extended for larger storage capacities. 
 
Our architecture works as follows: If a user requests one of the files stored in neighbors' caches in the RGG, neighbors will handle the request locally through D2D communication; otherwise, the BS should serve the request. Thus, to have D2D communication it is not sufficient that the distance between two users be less than $r(n)$; users should find their desired files locally in caches of their neighbors. A link between 
two users will be called potentially active if one requests a file that the other is caching. 
 Therefore, the probability of D2D collaboration opportunities depends on what is stored and requested by the users. 
 
The decision of what to store can be taken in a distributed or centralized way. 
A central control of the caching by the BS allows very efficient file-assignment to the users \cite{golrezaeiscalingglobcom}. However, if such control is not desired or the users are highly mobile, caching has to be optimized in a distributed way. The simple randomized caching policy we investigate makes each user choose which file to cache by sampling from a caching distribution. 
It is clear that popular files should be stored with a higher probability, but the question is that how much redundancy we want to have in our distributed cache.

We assume that all D2D links share the same time-frequency transmission resource within one cell area. This is possible since the distance between requesting user and user with the stored file will typically small. However, there should be no destructive interference of a transmission by others on an active D2D link. We assume that (given that node $u$ wants to transmit to node $v$) any transmission within range $r(n)$ from $v$ (the receiver) can introduce interference for the $u-v$ transmission. Thus, they cannot be activated simultaneously. 
This model is known as \emph{protocol model}; while it neglects important wireless propagation effects such as fading \cite{Molisch_book_2011}, it can provide fundamental insights and has been widely used in prior literature \cite{gupta2000capacity}.

\begin{figure}[ht]
\centering
\subfigure[]{
\centering \includegraphics[scale=0.4]{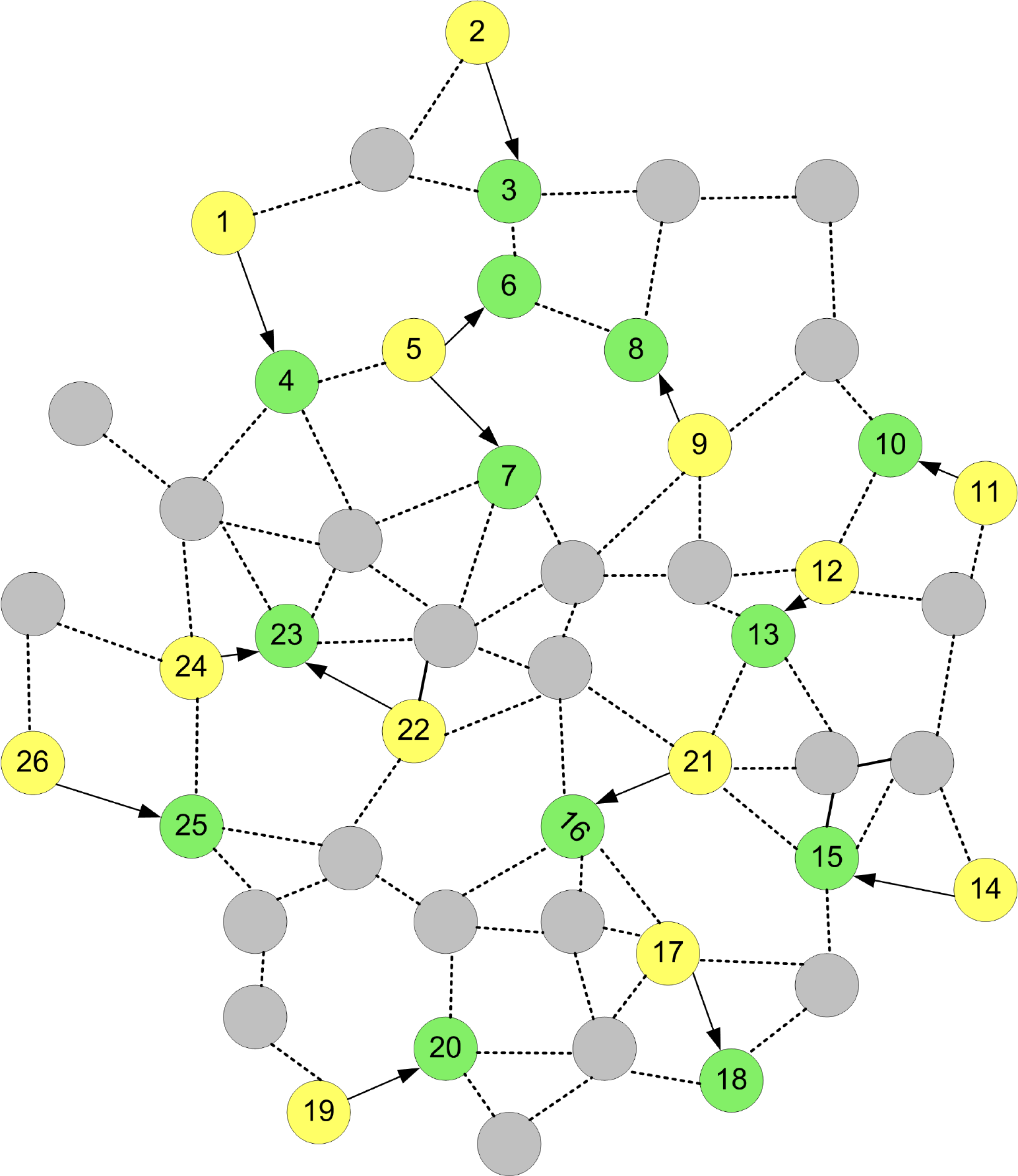}
\label{fig:subfig1}
}

\subfigure[ ]{
\centering \includegraphics[scale=0.6]{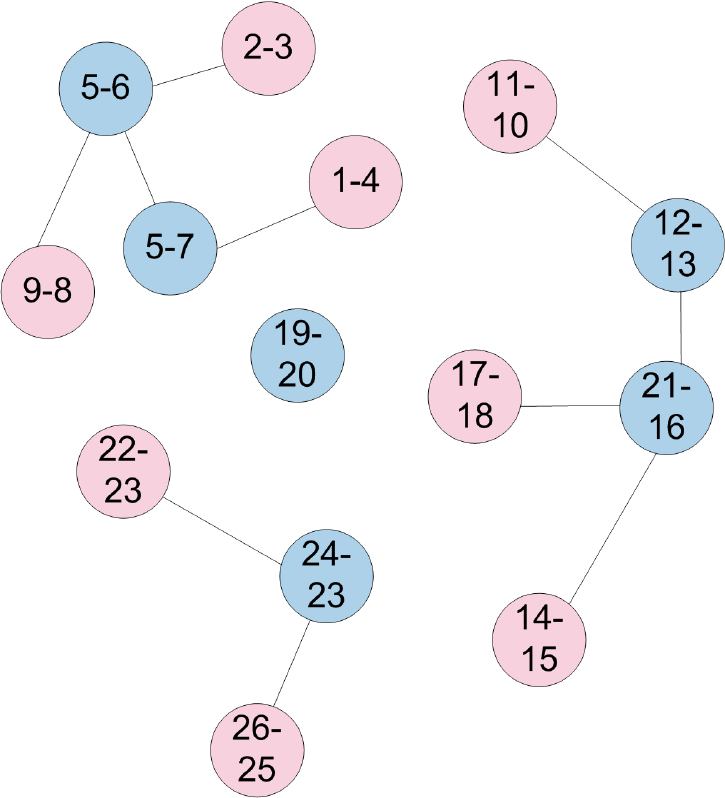}
\label{fig:subfig3}
}
\caption[Optional caption for list of figures]{a)~Random geometric graph, yellow and green nodes indicate receivers, transmitters in D2D links.  Gray nodes get their request files from the BS. Arrows show all possible D2D links.  b)~conflict graph based on Figure \ref{fig:subfig1} and one of maximum independent set of the conflict graph; pink vertices are those D2D links that can be activated simultaneously.}\label{conflict graph}
\end{figure}

To model interference given a storage configuration and user requests we start with all 
potential D2D collaboration links. Then, we construct the conflict graph as follows. We model any possible D2D link  between node $u$ as transmitter to node $v$ as a receiver with a vertex $u-v$ in the conflict graph. Then, we draw an edge between any two vertices (links) that create interference for each other according to the protocol model. Figure \ref{conflict graph} shows how the RGG is converted to the conflict graph. In Figure \ref{fig:subfig1}, receiver nodes are green and transmitter nodes are yellow. The nodes that should receive their desired files from the BS are gray. 
A set of D2D links is called active if they are potentially active and can be scheduled simultaneously, 
\textit{i.e.}, form an independent set in the conflict graph. The random variable counting the number of active D2D links under some policy is denoted by $L$. 

Figure~\ref{fig:subfig3} shows the conflict graph and one of maximum independent sets for the conflict graph. We can see that out of $14$ possible D2D links $9$ links can co-exist without interference. 
As is well known, determining the maximum independent set of an arbitrary graph is computationally intractable (NP complete~\cite{lawler1980generating}). 
Despite the difficulty of characterizing the number of interference-free active links, we can determine the best possible scaling law in our random ensemble. 

\section{Analysis}

\label{sec:analysis}

\subsection{Finding the optimal collaboration distance}
\label{Finding the optimal transmission radius}

We are interested in determining the best collaboration distance $r(n)$ and caching policy such that 
the expected number of active D2D links is maximized. 
Our optimization is based on balancing the following tension: The smaller the transmit power, the smaller the region in which a D2D communication creates interference. Therefore, more D2D pairs can be packed into the same area allowing higher frequency reuse.  On the other hand, a small transmit power might not be sufficient to reach a mobile that stores the desired file. Smaller power means smaller distance and hence smaller probability of collaboration opportunities. The optimum way to solve this problem would be to 
assign different transmit power to each node dynamically, to maximize the number of non-interfering collaborating pairs. However this approach would be intractable and non-practical.

Our approach is to enforce the same transmit power for all the users and show how to optimize it based on the content request statistics. Our analysis involves finding the best compromise between the number of possible parallel D2D links and the probability of finding the requested content.
Our results consist of two parts. In the first part (upper bound), we find the best achievable scaling for the expected number of active D2D links. In the second part (achievability), we determine an optimal caching policy and $r(n)$ to obtain the best scaling for the expected number of active links $E[L]$.

The best achievable scaling for the expected number of active D2D links depends on the extend of content reuse. Larger Zipf distribution exponents correspond to more redundancy in the user requests and a small number of files accounts for the majority of video traffic. Thus, the probability of finding requested files through D2D links increases by having access to few popular files via neighbors.

We separate the problem into two different regions depending on the Zipf exponent: $ \gamma_r>1$ and $\gamma_r<1$.
For each of these regimes, we find the best achievable scaling for $E[L]$ and the optimum asymptotic $r(n)$ denoted by $r_{opt}(n)$. 
We also show that a simple distributed caching policy with the properly chosen caching distribution 
has optimal scaling, \textit{i.e.}, matches the scaling behavior that any centralized caching policy could achieve\footnote{We use the standard Landau notation: $f(n)=O(g(n))$ and $f(n)=\Omega(g(n))$ respectively denote $|f(n)|\leq c_1 g(n)$ and $|f(n)|\geq c_2 g(n)$ for some constants $c_1,c_2$. $f(n)=\Theta(g(n))$, stands for $f(n)=O(g(n))$ and $f(n)=\Omega(g(n))$.  Little-o notation, \textit{i.e.,} $f(n) = o(g(n))$ is equivalent to $\lim_{n\rightarrow \infty} \frac{f(n)}{g(n)}=0$.}. 

Our first result is the following theorem:
\begin{theorem}\label{theorem_1}
If the Zipf exponent $\gamma_r >1 $,  
\begin{itemize}
\item[i)] \textbf{Upper bound:}
For any caching policy,
 $E[L]=O(n)$,
 \item[ii)] \textbf{Achievability:} Given that  $\sqrt{\frac{c_1}{n}}\leq r_{opt}(n)\leq \sqrt{\frac{c_2}{n}}$ and using a Zipf caching distribution with exponent $\gamma_c>1$ then $E[L]= \Theta(n)$.
\end{itemize}
\end{theorem}
The first part of the theorem \ref{theorem_1} is trivial since the number of active D2D links can at most scale linearly in the number of users.
The second part indicates that if we choose $r_{opt}(n)=\Theta(\sqrt{\frac{1}{n}})$ and $\gamma_c>1$, $E[L]$ can grow linearly with $n$. There is some simple intuition behind this result: We show that in this regime users are surrounded by a constant number of users in expectation. If the Zipf exponent $\gamma_c$ is greater than one, this suffices to show that the probability that they can find their desired files locally is a non-vanishing constant as $n$ grows. 
Our proof is provided in the Appendix \ref{proof1}.

For the low content reuse region $\gamma_r<1$, we obtain the following result:
\begin{theorem}\label{theorem2}
If $\gamma_r<1$, 
\begin{itemize}
\item[i)] \textbf{Upper bound:} For any caching policy, 
$E[L]=O(\frac{n}{m^{\eta}})$ where $\eta=\frac{1-\gamma_r}{2-\gamma_r}$,
 \item[ii)] \textbf{Achievability:} If $r_{opt}(n)=\Theta(\sqrt{\frac{m^{\eta+\epsilon}}{n}})$ and users cache files randomly and independently according to a Zipf distribution with exponent $\gamma_c$, for any exponent $\eta+\epsilon$,  there exists  $\gamma_c$ such that $E[L]=\Theta(\frac{n}{m^{\eta+\epsilon}})$  where $0<\epsilon<\frac{1}{6}$ and $\gamma_c$ is a solution to the following equation
 \[\frac{(1-\gamma_r)\gamma_c}{1-\gamma_r+\gamma_c}=\eta+\epsilon.\]
\end{itemize}
\end{theorem}

We show that when there is low content reuse, linear scaling in frequency re-use is not possible.
At a high level, in order to achieve the optimal scaling, on average a user should be surrounded by $\Theta(m^{\eta})$ users. Comparing with the first region where $\gamma_r> 1$, we can conclude that when there is less redundancy, users have to see more users in the neighborhood to find their desired files locally. Due to space constraints we omit this proof.

\section{Discussion and Conclusions}

\label{sec:discussion}

The study of scaling laws of the capacity of wireless networks has received significant attention since the pioneering work by Gupta and Kumar~\cite{gupta2000capacity} (\textit{e.g.} see \cite{Tse,grossglauser2001mobility,franceschetti2009capacity}).
 The first result was pessimistic: if $n$ nodes are trying to communicate (say by forming $n/2$ pairs), since the typical distance in a 2D random network will involve  roughly $\Theta(\sqrt{n})$ hops, the throughput per node must vanish, approximately scaling as $1/\sqrt{n}$. There are, of course, sophisticated arguments performing rigorous analysis that sharpens the bounds and numerous interesting model extensions. One that is particularly relevant to this project is the work by Grossglauser and Tse~\cite{grossglauser2001mobility} that showed that if the nodes have infinite storage capacity, full mobility and 
there is no concern about delay, constant (non-vanishing) throughput per node can be sustained as the network scales. 

Despite the significant amount of work on ad hoc networks, there has been very little work on 
file sharing and content distribution over wireless (\cite{femtocaching,chen2011file}) beyond the multiple unicast traffic patters introduced in~\cite{gupta2000capacity}. 
Our result shows that if there is sufficient content reuse, non-vanishing throughput per node can be achieved, even with constant storage and delay. In our recent work\cite{ICC_Workshop} we empirically analyzed the optimal collaboration distance for fixed number of users. 

On a more technical note, the most surprising result is perhaps the fact that in Theorem 2, a simple distributed policy can match the optimal scaling behavior $E[L]=O(\frac{n}{m^{\eta}})$. Further, for both regimes, the distributed caching policy exponent $\gamma_c$ should not match the request Zipf exponent $\gamma_r$, something that we found quite counter intuitive. 

Overall, even if linear frequency re-use is not possible, we expect the scaling of the library $m$ to be quite small (typically logarithmic) in the number of users $n$. In this case we obtain near-linear (up to logarithmic factors) growth in the number of D2D links for the full spectrum of Zipf exponents.  Our results are encouraging and show that distributed caching can enable collaboration and mitigate wireless content delivery problems.

\appendices
\section{Proof of Theorem 1}\label{proof1}
The first part of the theorem is easy to see since the number of D2D links cannot exceed the number of users.

\begin{figure}
\centering
\subfigure[]{
\centering \includegraphics[width=4.5cm]{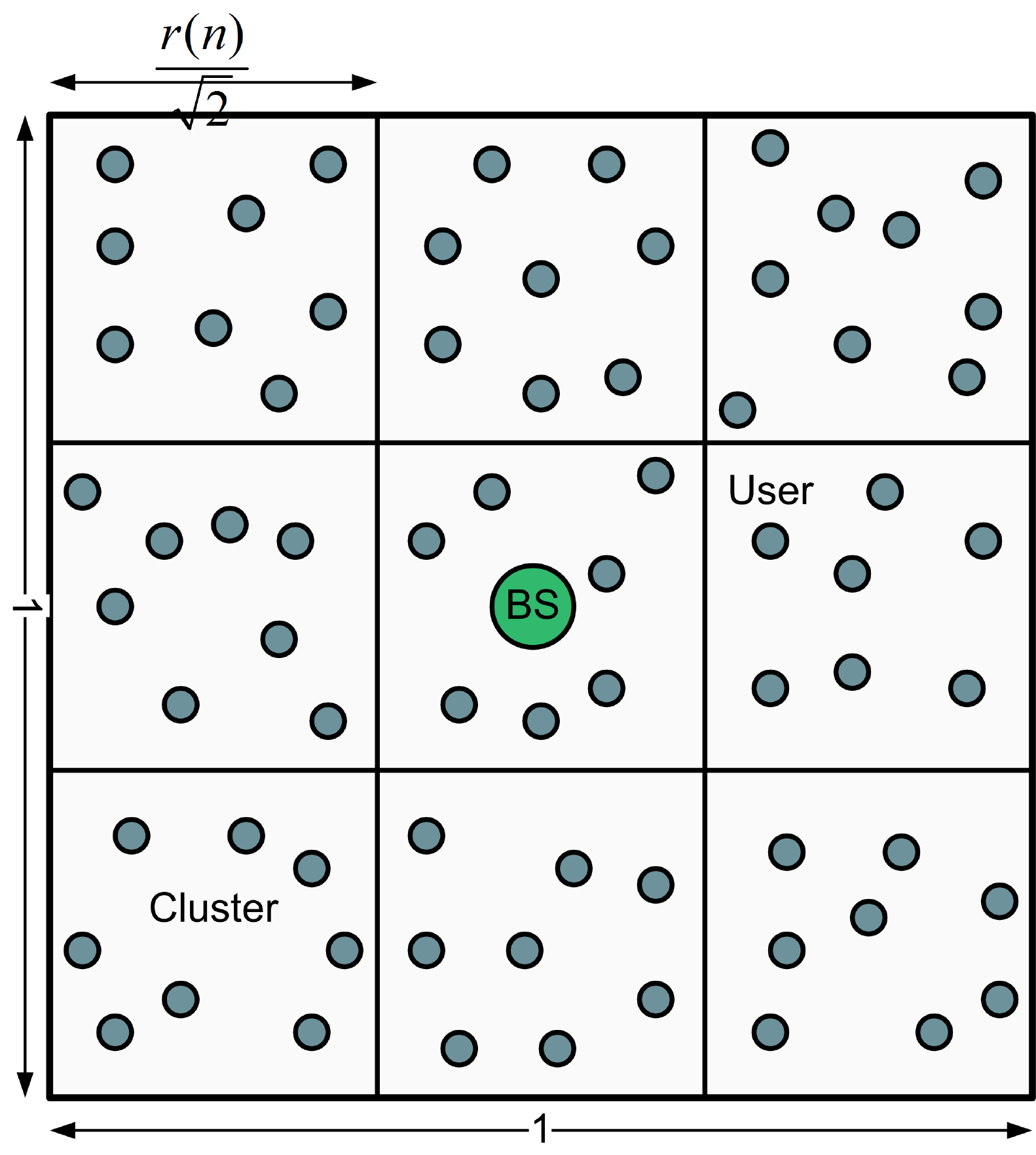}
\label{layout}
}

\subfigure[ ]{
\centering \includegraphics[width=4.5cm]{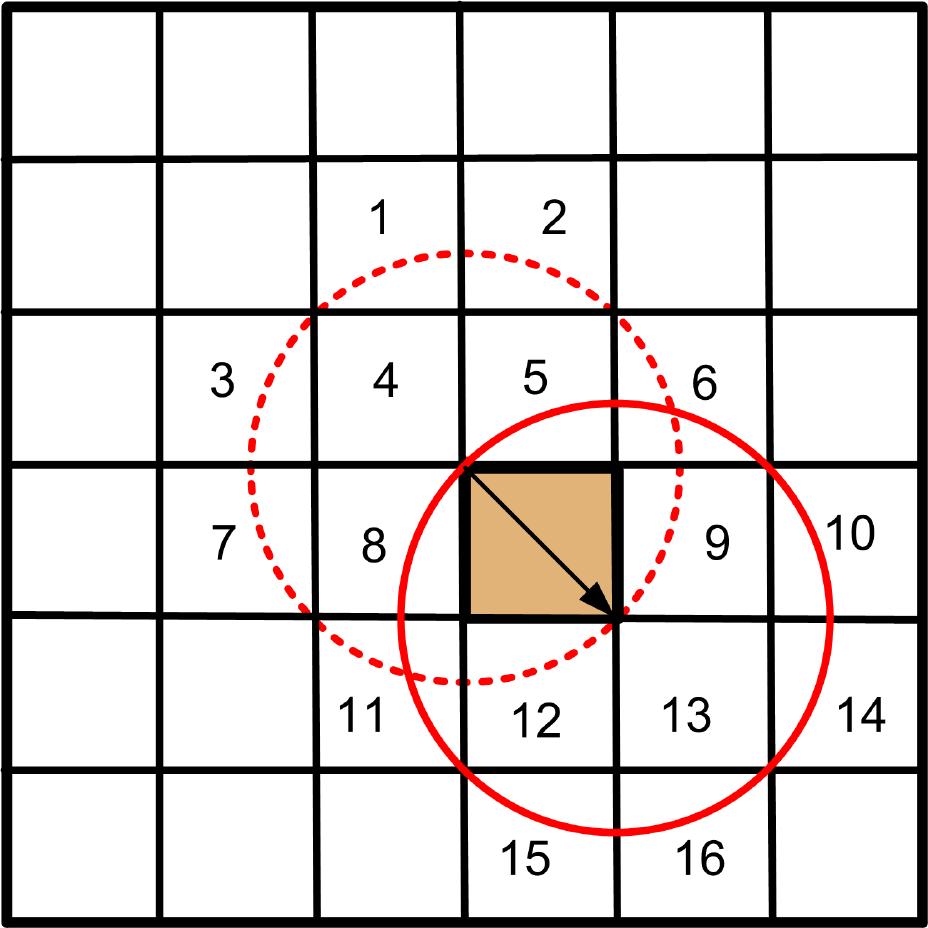}
\label{interfer}
}
\caption{a)~Dividing cell into virtual clusters. b)~In the worst case, a good cluster can block at most $16$ clusters. In the dashed circle, receiving is not possible and in the solid circle, transmission is not allowed.}
\end{figure}

For the second part of theorem \ref{theorem_1}, we divide the cell into $\frac{2}{r(n)^2}$ virtual square clusters. 
 Figure \ref{layout} shows the virtual clusters in the cell. The cell side is normalized to $1$ and the side of each cluster is equal to $\frac{r(n)}{\sqrt{2}}$. Thus, all users within a cluster can communicate with each other. Based on our interference model, in each cluster only one link can be activated. 
  Thus, to prove the theorem, it is enough to show that in a constant fraction of virtual clusters, there are active D2D links that do not introduce interference to each other. This is because $r(n)=\Theta(\sqrt{\frac{1}{n}})$ and there are $\Theta(n)$ virtual clusters in the cell. When there is an active D2D link within a cluster, we call the cluster \emph{good}. But not all good clusters can be activated simultaneously. One good cluster can at most block $16$ clusters (see Figure \ref{interfer}). The maximum interference happens when a user in the corner of a cluster transmits a file to a user in the opposite corner. So, we have
 $E[L]\ge \frac{E[G]}{17}$
 where $E[G]$ is the expected number of good clusters.  Since we want to find the lower bound for $E[L]$, we can limit users to communicate with users in  virtual clusters they belong to. 
 Therefore, we have
 \begin{align}
 \nonumber E[G]
 &\geq \frac{2}{r(n)^2}\sum_{k=0}^{n}{\Pr[\text{good}|k]\Pr[K=k]},
 \end{align}
 where $\frac{2}{r(n)^2}$ is the total number of virtual clusters. $K$ is the number of users in the cluster, which is a binomial random variable with $n$ trials and probability of $\frac{r(n)^2}{2}$, \textit{i.e.,} $K=B(n,\frac{r(n)^2}{2})$.  $\Pr[K=k]$ is the probability that there are $k$ users in the cluster and $\Pr[\text{good}|k]$ is the probability that the cluster is good conditioned on $k$. The probability that a cluster is good depends on what users cache. Therefore,
\begin{align}\label{EG111}
\nonumber  E[G]&\geq \frac{2}{r(n)^2}\sum_{k=0}^{n}{\Pr[K=k]}\\
&\times  \sum_{ \big\{{\bf{u}}\, \big| |{\bf{u}}|=k\big\}} \Pr[\text{good}|{\bf{u}},k]\Pr[{\bf{U}}={\bf{u}}],
 \end{align} 
 where $\bf{U}$ is a random vector of stored files by users in the cluster. $\bf{u}$ is a realization of $\bf{U}$ and $|\bf{u}|$ denotes the length of vector $\bf u$. 
The  $i$th element of $\bf u$ denoted by ${\bf u_i}\in \{1,2,3,\ldots, m\}$ indicates what user $i$ in the cluster stores.

For each $\bf{u}$, we define a value:
\[v(\bf u)=\sum_{i\in \tilde{ \bf u}} f_i,\]
where $\tilde{ \bf u}=\cup_{j=1}^{|\bf u|}\bf u_j$ and $\cup$ is the union operation. Actually $v({\bf{u}})$ is the sum of popularities of the union of files in $\bf{u}$.
The cluster is considered to be good if at least a user $i$ in the cluster requests one of the files in $\tilde{ \bf u}-\{\bf{u}_i\}$. 
 Note the possibility of {\em self-requests}, \textit{i.e.,} a user might find the file it requests in its own cache; in this case clearly no D2D communication will be activated by this user. Accounting for these self-requests,
 the probability that user $i$ finds its request files locally within the cluster is $(v({\bf{u}})-f_{\bf{u}_i})$. Thus, we obtain:
\begin{align}\label{pr[good]111}
 \Pr[\text{good}|{\bf{u}},k]
&\ge 1-(1-(v({\bf{u}})-\max_{i} f_{\bf{u}_i}))^k.
\end{align}
 Let us only consider cases where at least one user in the cluster caches file $1$ (the most popular file). Then, from (\ref{EG111}) and (\ref{pr[good]111}),  the following lower bound is achieved:
 \begin{align}
 \nonumber E[G]&\ge \frac{2}{r(n)^2}\sum_{k=1}^{n}{\Pr[K=k]}\\
 &\times \sum_{\bf{u}\in x}1-(1-(v({\bf{u}})-f_{1}))^k \Pr[{\bf{U}}={\bf{u}}].\label{E[G]22}
 \end{align}
 where ${\bf{x}}={ \big\{{\bf{u}}\, \big| |{\bf{u}}|=k\, \, \text{and} \,\,1\in {\bf u}\big \}} $.
 Let us further define a random variable $V$ which is sum of popularities of the union of files stored by users in the cluster. Then, in equation  (\ref{E[G]22}), we can take the expectation with respect to $V$, \textit{i.e.,} 
  \begin{align}
 \nonumber E[G]&\ge  \frac{2}{r(n)^2}\sum_{k=1}^{n}{\Pr[K=k] E_V[1-(1-(V-f_1))^k|A_1^k]}\\
  \nonumber& \ge \frac{2}{r(n)^2}\sum_{k=1}^{n}{\Pr[K=k] E_V[(V-f_1)|A_1^k]},
 \end{align}
 where $A_{1}^k$ is the event that at least one of $k$ users in the cluster caches file $1$ and $E_V[.]$ is the expectation with respect to $V$. Let $A_{1,h}^k$ for $1 \leq h\leq  k$ denote the event that $h$ users out of $k$ users in the cluster cache file $1$. Then, we get:
 \begin{align}\label{E[G]2}
 \nonumber E[G]& \ge \frac{2}{r(n)^2}\sum_{k=1}^{n}{\Pr[K=k] \sum_{h=1}^{k}E_V[(V-f_1)|A_{1,h}^k] }\\
 &\times 
 \left( \begin{array}{c}
k  \\
h \end{array} \right) (p_1)^h
(1-p_1)^{k-h},
 \end{align}
 where $p_j$ represents the probability that file $j$ is cached by a user based on Zipf distribution with exponent $\gamma_c$.
 To calculate $E_V[(V-f_1)|A_{1,h}^k]$, we define an indicator function ${\bf 1}_{j}$ for each file $j\ge 2$.  ${\bf 1}_{j}$ is equal to 1 if at least one user in the cluster stores file $j$. Hence,
 \begin {align}
 \nonumber E_V[(V-f_1)|A_{1,h}^k]&=E[\sum_{j=2}^m f_j {\bf 1}_{j}|A_{1,h}^k]\\
\nonumber & =\sum_{j=2}^m f_j (1-(1-p_j)^{k-h}).
 \end{align}
 Substituting $E_V[(V-f_1)|A_{1,h}^k]$ in (\ref{E[G]2}) and limiting the interval of $k$, we can obtain:
 \begin{align}
 \nonumber &E[G]
  \ge \frac{2}{r(n)^2}\sum_{k\in I}\Pr[K=k]\times \\
 & \sum_{h=1}^k  \sum_{j=2}^{m} f_j (1-(1-p_j)^{k-h}) 
 \left( \begin{array}{c}
k  \\
h \end{array} \right) (p_1)^h
(1-p_1)^{k-h},
 \label{E[G]theorem1}
 \end{align} 
 where $0<\delta<1$ and $I= [nr(n)^2(1-\delta)/2,nr(n)^2(1+\delta)/2]$. Define $k^*\in I$ such that it minimizes the expression in the last line of  (\ref{E[G]theorem1}).
 Considering that $r(n)=\Theta(\sqrt{\frac{1}{n}})$, $k^*$ is $\Theta(1)$. Then from (\ref{E[G]theorem1}), we have:
 \begin{align}
 \nonumber &E[G] \ge \frac{2}{r(n)^2}\Pr[k\in I] \sum_{h=1}^{k^*}  \sum_{j=2}^m f_j(1-(1-p_j)^{k^*-h}) \\ 
 \label{before_chernoff1}&
 \times \left( \begin{array}{c}
k^*  \\
h \end{array} \right) (p_1)^h
(1-p_1)^{k^*-h}\\
&\nonumber  \ge   \frac{2}{r(n)^2} (1-2e^{-nr(n)^2\delta^2/6}) \sum_{h=k^*p_1(1-\delta_1)}^{k^*p_1(1+\delta_1)}\Big[ \left( \begin{array}{c}
k^*  \\
h \end{array} \right) \\ 
&\times \sum_{j=2}^m f_j(1-(1-p_j)^{k^*-h}) 
  (p_1)^h
(1-p_1)^{k^*-h} \Big],\label{chernoff1}
 \end{align} 
where $0<\delta_1<1$. 
We apply the Chernoff bound in (\ref{before_chernoff1}) to derive (\ref{chernoff1}) \cite{chernoff1952measure}. Since the exponent $nr(n)^2\delta^2/6$ is $\Theta(1)$, we can select the constant $c_1$ such that the term $(1-2e^{-nr(n)^2\delta^2/6})$ becomes positive.

Let us define $h^*\in [k^*p_1(1-\delta_1),k^* p_1(1+\delta_1)]$ such that it minimizes the expression in the last line of  (\ref{chernoff1}).
From (\ref{zipf}) and lemma \ref{scaling}, $p_1$ is $\Theta(1)$ and as a result, $h^*$ is also $\Theta(1)$.
Using the Chernoff bound in (\ref{chernoff1}), we get:
\begin{align} \nonumber 
 E&[G]\ge \frac{2}{r(n)^2} (1-2e^{-nr(n)^2\delta^2/6}) (1-2e^{-k^*p_1\delta_1^2/3})\\ 
&\times \left( \begin{array}{c}
k^*  \\
h^* \end{array} \right) (p_1)^{h^*}
(1-p_1)^{k^*-h^*}\sum_{j=2}^m f_j(1-(1-p_j)^{k^*-h^*}).\label{summation}
\end{align}
   $k^*-h^*$  should be greater than $1$ which results in a constant lower bound for $c_1$. The second exponent, \textit{i.e.,} $k^*p_1\delta_1^2/3$ is $\Theta(1)$. The term  $(1-2e^{-k^*p_1\delta_1^2/3})$ is a positive constant if  $c_1\ge \frac{3\ln 2 \zeta(\gamma_c)}{\delta_1^2(1-\delta)}$, where $\zeta(\gamma)=\sum\limits_{j=1}{\frac{1}{j^{\gamma}}}$ is the Riemann zeta function \cite{Riemann}. Further, the summation in (\ref{summation}) satisfies 
\[\sum_{j=2}^m f_j(1-(1-p_j)^{k^*-h^*}) >\sum_{j=2}^m f_jp_j.\]
To show that $E[G]$ scales linearly with $n$, the term  
$\sum_{j=2}^m f_jp_j$ should not be vanishing as $n$ goes to infinity.  
It can been shown that 
if $\gamma_r,\gamma_c>1$, 
$\sum_{j=2}^{m}f_j p_j=\Theta(1)$ (see lemma \ref{scaling}).

\begin{lemma} \label{scaling}
 If $ \gamma>1$, $a=o(b)$, and $a=\Theta(1)$, then 
$H( \gamma,a,b)=\Theta(1)$ and 
$\sum_{j=a}^{b}f_j p_j=\Theta(1)$
where $H(\gamma,a,b)=\sum\limits_{j=a}^{b}\frac{1}{{{i^\gamma }}}$.
\end{lemma}
The proof is omitted due to lack of space.

\bibliographystyle{IEEEtran}
\bibliography{ref_gupta.bib}

\end{document}